# Time-resolved measure technique for electron beam envelope basing on synchronous framing and streaking principle*

Jiang Xiaoguo，Wang Yuan，Yang Zhiyong，Zhang Huang，Wang Yi，Wei Tao
（ Institute of Fluid Physics，China Academy of Engineering Physics，Mianyang，621900，China ）

**Abstract:** The time-resolved electron beam envelope parameters including sectional distribution and position are important and necessary for the study of beam transmission characteristics in the magnetic field and verifying the magnetic field setup rationality. One kind of high time-resolved beam envelope measurement system has developed recently. It is mainly constituted of high framing camera and streak camera. It can obtain 3 panoramic images of the beam and the time continuous information of the given beam cross section at one time. The recently obtained data has proved that several fast vibration of beam envelope along the diameter direction occur during the rising edge and the falling edge of the electron beam. The vibration period is about several nanoseconds. The effect of magnetic field on the electron beam is also observed and verified. The beam debug experiments have proved that the existing beam transmission design is reasonable and viable. The beam envelope measurement system will establish a good foundation for beam physics research.

**Key words:** high speed framing camera；streak camera；beam envelope；time-resolved measurement；

PACS：41.85.Ja；41.85.Ew；29.27.Eg

The linear induction accelerator(called LIA) is one kind of accelerators which is used to accelerate electron beam of several kilampere current. The pulsed electron beam of high intense current must be well restricted by a strong magnetic field into the center area of accelerating cavity with several centimeters in diameter and the electron beam should be transmitted without any losing as possible from the injector to the exit of the accelerator. The transmitting distance is about several decameters. This is a great challenge for high intense current electron beam transmit. The most important work for transmitting beam in the injector is how to get a best matched magnetic field configure. The mismatched magnetic field configure at the stage of injector and the first accelerating cavity will bring serious influence on the following beam transmitting because it is situated at the source stage.The focus force or restriction ability on the electron beam will obviously varied with the electron energy and the emission status of the electron from the thermionic cathode at the injector stage[1, 2]. This results that electrons with different energy move ahead along different tracks inside the cavity and that the beam envelope vary markedly under some certain mismatched magnetic field configure. The primary theoretic simulation shows that the electron beam envelope shrinks and enlarges tempestuously during the rising and falling period of the electron beam pulse. There are usually about 3 ~ 5 periods during about 20 nanoseconds. Because of the tempestuous beam envelope vibration, some electron may run into the cavity wall and should be lost. Electron beam with better character may be obtained if losing voluntarily the unwanted electron current at some certain position by the way of adjusting magnetic field configure. But the lost electron current may be likely to bring some ill effect to the following beam transmission. On the other hand, the electron beam envelope may be well restricted under some matched magnetic field configure and can be transported to the next stage without electron lost or electron beam envelope vibration.The time-resolved measurement technology especially for electron beam envelope is the key technology to debug the electron beam. But it is very difficult to obtain about 5 periods of vibrations during about 20 nanoseconds at the edge of beam pulse. The time-resolved measurement ability and attainable image frames are critical limit to the measurement system for obtaining the detail information about the electron beam envelope tempestuous changing. Basing on the developed time-resolved measurement system[3,4,5,6,7] and design principle in hand, one set of compositive measurement system for electron beam envelope is built up by the way of combining high speed framing camera and streak camera. The several framing panorama of the electron beam envelope at different time and the profile image of the center slit of the electron beam in the whole pulse duration can be obtained at one time. The framing images are captured by high speed framing camera with about several nanoseconds shutter time . The interval time between the framing images can reach 3ns ~ 5ns. The profile image is captured by the streak camera. The time-resolved ability for the profile image can reach sub-nanosecond. This measurement ability can meet the above mentioned demand of obtaining the tempestuous changing of the beam envelope at the rising and falling edge of the pulse.   This measure technology achieves successful applications in the beam debugging and beam transmission at the injector stage. It has provided plentiful experiment data including intuitionistic and exact image for the debugging work magnetic field configure at the injector stage. It has established a good foundation for obtaining the beam vibration. It provides experimental measurement ability and validation ability for the best matched magnetic field configure debug and the study of the beam transmission theory.

* Supported by National Natural Science Foundation of China (10675104，11375162)
1) E-mail: j_xg_caep@sina.com

# 1    The structure of the time-resolved measurement system for the beam envelope basing on framing and streak

Figure 1 shows the system structure of the time-resolved beam envelope measurement system. It is made up of optical transition radiation(called OTR ) target[8], light beam splitting component, high speed framing camera and high speed streak camera.The highest time-resolved ability of this system is mainly restricted by the camera speed. The minimum shutter time is about 2ns and the framing camera can capture 3 frame images at one time. The streak camera can provided with about 2ps temporal system resolution at fastest speed and it usually work at sweep speed of about 0.2ns/point. The time-resolved ability of this united system can meet the framing and streak imaging demands with enough speed for the pulsed electron beam with duration of about 100ns. The BPM in figure 1 is beam position monitor which is used to measure the beam current and the beam center position. The BPM is located at 500mm upriver from the OTR target.

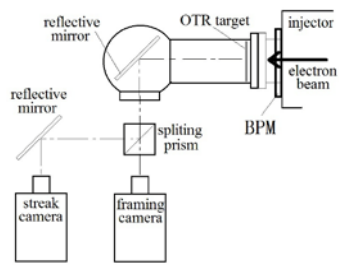
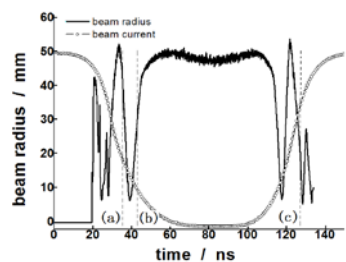

Fig.1    Layout of the time-resolved electron beam measurement        Fig.2    Simulative beam envelope changed with time

# 2 The related study on electron beam envelope and the time-resolved measurement

The electron beam shot from the thermionic cathode features wider energy spectrum, some transverse momenta. This will bring some problems to the electron beam transport and acceleration and these problems must be solved at first at injector stage. In the injector , the electron beam is shot form the thermionic cathode first and run into the anode stage and then is transported. At this stage, the electron beam is expected to be transported lossless as possible. The beam envelope is also expected not to change tempestuously as possible and the beam can be stably   transited to the acceleration stage. This can restrain beam emittance growth during transmission and acceleration. How to configure the magnetic field at the injector stage is an important research work. The matching grade of magnetic field will affect the electron beam transmission efficiency.

The theoretic simulation work is carried out according to the actual structure of the diode in the injector. The simulative work is aimed at the temporal continuity of the beam envelope. The measurement of beam envelope is also accomplished for the relative conditions. A series of useful results is obtained. It plays an important role in analyzing the effect of magnetic field and validating simulative program.

The typical simulation results of beam envelope profile continuous changed with time is shown in the figure 2. In order to be associated with the beam current waveform in time, the amplitude is normalized and the position is offset. The simulation is aimed at the beam envelope changing during about 150ns at the position of 4900mm from the cathode surface. The simulation starts at 0ns. The simulation results show several vibrations during the leading edge and the following edge of electron beam pulse. But these phenomena aren't observed in experimentation ago. The simulation results show that there are obvious difference between different magnetic field configure. The beam diameter can be large or small during the pulse. The beam envelope will change in different form. These will produce important and some uncertain influence. It very important that the beam envelope change rule would be found out.

Figure 3 shows some typical framing and streak images of electron beam at the exit of the injector. The image contrast and image gray are adjusted to the optimized visual impression. Image (a),(b)and (c) are framing images. Image (a) and (b) are captured during the leading edge of beam pulse. The shutter opening monitor shows that the shot time of image (a) is 36ns and the shot time of image (b) is 43ns. Image (c) is captured during the following edge of beam pulse. The shot time of image (c) is 127ns. All shutter time of above images are 5ns. The shot time associated with the beam current waveform is shown in figure 2. The beam envelope images can be captured during the leading edge or the following edge of the short duration of electron beam pulse. It is very difficult to obtain the entire course of the beam envelope vibration because of the following main factors. 1, the total images captured by the framing camera is only 3 frames. It is not enough to observe the whole course. 2,the beam envelope vibration is a really very rapid course to the framing camera. If a wider shutter time is adopted to capture the vibration image, the captured image is really an integral image of some duration compared with the vibration. This is one factor that the beam image appears to be layered some times. 3, The enough image gray can't be obtained when a shorter shutter time is adopted to capture the rapid image. This will result in a low contrast image and not well distinguished image. 4, It is very difficult to synchronize the shutter with the beam pulse edge in ns level because the trigger synchronization precision of the framing camera is about 5ns in the LIA environment. The streak camera takes on preferable advantages of continuously rapid sweep and low synchronization accuracy demand for trigger. Image (d) in figure 3 is the streak image of the electron beam profile during the whole beam pulse . The streak camera images at the middle position of the beam profile by the way of adjusting the slit. The image captured shows the diameter information about the beam profile at the middle. This streak image shows indeed that there are several beam vibration during the leading edge and the following edge of the beam pulse. The beam diameter during pulse middle also change.

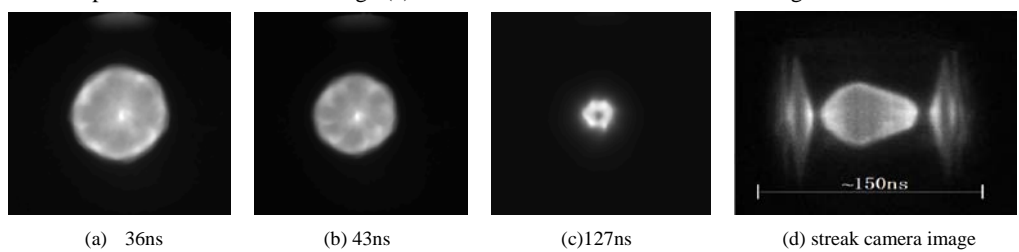

(a)　36ns　　　　　(b) 43ns　　　　　(c)127ns　　　　　(d) streak camera image

Fig.3　Typical time-resolved framing images and streak camera image of electron beam from the injector

If some electron run into the cavity wall at the injector , some ill effect to the following beam acceleration and transmission would occur. Some validating experimentations are designed to study this phenomena. There is only one BPM in the inner cavity at the injector. The BPM is made of 48 resistances and these resistances are ranged in parallel connection along the radial direction. There may be some given spatial distribution at this BPM position. The correlative beam profile with the BPM ring structure may be produced if the electron run into the cavity wall at BPM position. We can judge this status. If electron run into the cavity wall at some other position, we can't judge well because of the ring wall without any characters.A set of magnetic field configure, which will cause electron run into the wall at BPM ,is obtained by the PIC simulation program and some experiments are carried out. Figure 4 shows the typical beam envelope images captured. Image (a) and (b) are two images during the leading edge captured by framing camera. Image (a) is captured about 8ns early before image (b). They may be captured between 20ns and 30ns course in figure 2. Image (b) may show the last vibration status before the flat top. Image (c) shows the beam envelope status during the flat top. All shutter time for above images are 5ns. These experiment phenomena can be reproduced steadily.The beam current measured by the BPM at exit of the injector is lost obviously compared with the emission current at surface of the thermionic cathode. The beam current waveform shows that the loss during the leading edge and the following edge may be large especially and the loss during the flat top is small. This status indicates that there is a lot of electron collided with the wall. The 48 toothed fringe and 48 spoke in the image (a),(b) and (c) may prove that the colliding location is at the BPM. The electron beam shape will not change obviously from the BPM to the OTR target because of the very close position. The beam profile don't diverge or focus instantaneously and keep the shape it has at the BPM on the whole. Although the three images show the course of the electron beam profile evolvement , how it happen is not clear because of the rapid vibration. The mechanism of forming this kind of images is an other main topic which will be researched. The diameter of the electron beam at the leading edge of the pulse is go beyond the inner wall of the cavity estimated form the streak image and the electron beam is then restricted by the wall. The electron must collide with the wall at the time. Under this magnetic field configure, the beam diameter during the flat top section will reduce gradually and slowly. This is shown in figure 4 (d).

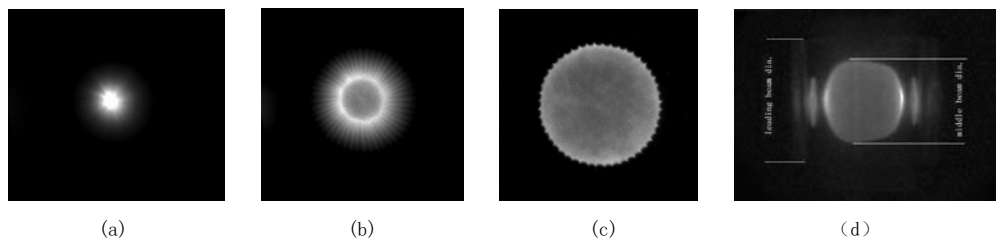

(a)  (b)  (c)  (d)

Fig.4  Typical images produced when electron beam collide with the inner wall at the BPM

The above simulation work establish a good foundation of controlling the beam envelope. The other magnetic field configure is simulated again in order to transport the electron beam without lossness to the back . This magnetic field configure will insure that the electron will be restricted in the center area of accelerating cavity to run forth without lossness and that the electron beam profile diameter is smaller than the cavity diameter. In this case, the beam envelope will change slowly without any tempestuous. The typical image captured under this case is shown in figure 5. Image (a) is framing image captured at the flat top section of the beam pulse with 5ns shutter time. Image (b) is streak image captured by streak camera with 200ns sweep time. This streak image shows some decrease in amplitude and intensity of the beam envelope vibration during the leading edge and the following edge. We also observe that the beam diameter during the flat top section of the pulse keep stabilization and that the beam profile has good uniformity. The results have proved that the magnetic field configure is relatively reasonable. The later beam transport work and focusing beam work also show the improved effect.

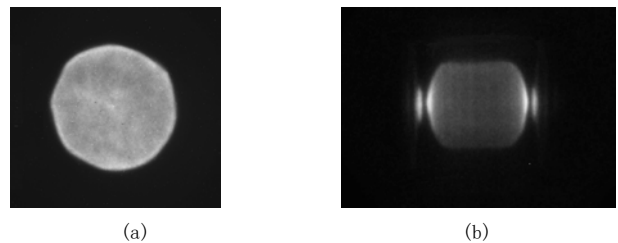

(a)  (b)

Fig.5  verification results images when electron beam transmission without loss through the injector

More sets of magnetic field configure at the injector are also researched in order to compare the theoretic simulative results with the actual results. This work achieves expectant purpose as a whole. The figure 6 show a series of actual images and the theoretic simulative results under three magnetic field configures. The curve is simulative results of beam envelope vibration and the picture is its actual streak images. In general, the simulative results match the actual images more at the main section.

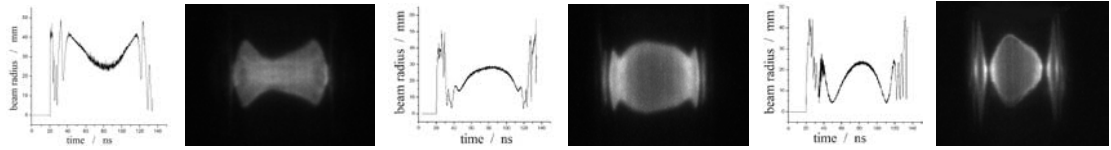

Fig.6　Several simulated beam envelope and the relevant practical streak images under different magnetic field

## 3 Discussion

The actual images especially captured by the streak camera have shown that there are several beam envelope vibrations during the leading and following edge of the pulse and the vibrations are tempestuous some time. The vibration amplitude and duration are different from each other even if compared within the same pulse. The theoretically wanted purpose usually can be achieved to a great extend by adjusting the magnetic field configure according to the simulative results and the actual results also match the simulative results. This work has proved that we have basically possessed simulation ability of intense current pulsed electron beam transmission at the injector stage. This ability provide very helpful assistance in debugging work of pulsed electron beam of several kilamperes. This may be the most fruitful research result derived from this paper.But there is some difference in detail between the actual image and the simulative results if observed carefully. The vibration number observed may be different with simulative results and the beam envelope shape is some different from the actual captured images. There may be two main cause. First of all, there may be some difference between the input electron beam parameters for the theoretic simulation and the actual value. The difference may be notable sometimes. The electron beam parameters include beam emittance, electron beam incidence angle, beam energy, beam current, current waveform and so on. How to choose these parameter value is a key work in theoretic simulation. The influence of beam emittance and electron beam incidence angle may be remarkable. But it is very difficult for us to get an exact parameters value or to measure it. The parameters are choose by some experience and this results that it is not far best matching between the theoretic simulation and the actual value at present. On the other hand, the image gray of the front part on the leading edge or the back part on the following edge is too low to be seen. This results that the outline of the beam envelope can be faintly seen and that the vibration number can't be clear distinguished. The difference of electron energy would induce different vibrating period and different beam envelope diameter even if under the same magnetic field. The beam envelope with different diameter and different vibrating period would overlap with each other and the outline of the beam profile image will be confused. The electron density here would reduce and the image gray become lower. The width of the slit at the input window of the streak camera may be set to a wider to obtain enough image gray. This reduce the time-resolved ability further and this equally reduce the spatial resolution along the time axis in the image. It increase some difficulty to distinguish the rapid beam envelope vibration. But several beam envelope vibration during the leading or following edge can be captured under specific conditions shown in figure 7 if some means are adopted. The main method is increasing sweep speed and only image during one part of the duration is captured. Although the entire image of duration can't be obtained, the detail information of beam envelope at the given duration can be obtained and the information is much clear. For example, some electron cloud outer can be seen in the image (b) in figure 7.

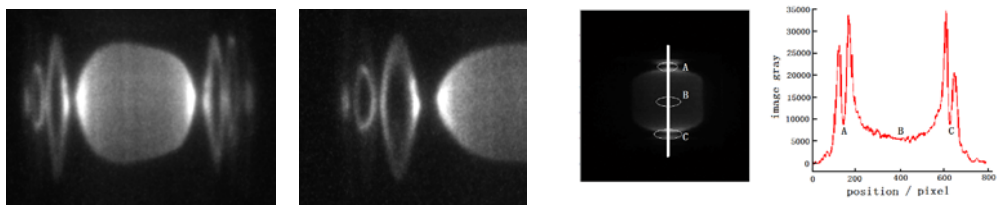

|  (a) | (b) | (a) original image | (b) profile data |

Fig.7  Streak images especially for the leading edge      Fig.8  Analysis on the time continuity for beam envelope

The other confused phenomenon in the streak image is the part between the last vibration and the flat top, shown as A and C in figure 8(a). It seems some discontinuity of electron beam at these part in the image. But the image gray at these part is really at an upper level if we get the image gray data and it may be greater than the image gray at the flat top section such as B in the image, shown as figure 8(b). The factual image gray prove that the electron beam is temporal continuity. The image discontinuity in vision may be caused by the remarkable differential contrast because the image gray difference is great.

## 4 Conclusion

The electron beam envelope measurement system is developed latest. It is made of newest developed high speed framing camera and streak camera. It can capture time-resolved framing images and can capture the streak image at the same time. These images are complementary and can explain some phenomena more comprehensively. The successful development of this electron beam envelope measurement system provide a strong and useful means to study pulsed intense current electron beam and its transmission rule. The recently obtained data from experimentation has proved that several fast vibration of beam envelope along the diameter direction do occur during the rising edge and the falling edge of the electron beam and that the vibration period is about several nanoseconds. The effect rule of magnetic field configure on electron beam transmission is also validated to some extent. The measurement work also prove that the existing beam transmission design is reasonable and viable. The theoretic simulation has practically leading effect on the electron beam debugging work at present. The far-reaching significance is that this system provide us with the ability of validating the theoretic simulation results and the beam physics research.